%% file: main.tex
\documentclass[letterpaper]{article}
\usepackage[preprint]{aaai2027}
\usepackage[hyphens]{url}
\usepackage{graphicx}
\usepackage[numbers,sort&compress]{natbib}
\usepackage{caption}
\usepackage{amsmath}
\usepackage{amssymb}
\usepackage{booktabs}
\usepackage{multirow}
\usepackage{array}
\usepackage{algorithm}
\usepackage{algorithmic}
\usepackage[
  colorlinks=true,
  linkcolor=blue,
  citecolor=blue,
  urlcolor=blue,
  bookmarks=true,
  bookmarksopen=true,
  pdfstartview=FitH
]{hyperref}

\makeatletter
\newcommand{\NeedColumnSpace}[1]{%
  \par
  \ifdim\pagegoal<\maxdimen
    \dimen@=\pagegoal
    \advance\dimen@ by -\pagetotal
    \ifdim\dimen@<#1\relax\newpage\fi
  \fi
}
\makeatother

\title{Trigger the Straggler: Load Hijack on Mixture-of-Experts LLMs}

\author{
Rui Zhang, Wenbo Jiang\thanks{Corresponding author (Wenbo Jiang): \href{mailto:wenbo_jiang@uestc.edu.cn}{\nolinkurl{wenbo_jiang@uestc.edu.cn}}; first author (Rui Zhang): \href{mailto:ruizhangsec@163.com}{\nolinkurl{ruizhangsec@163.com}}.}, Hongwei Li, Zihan Wang, Rui Zhang,\\
Chaoshun Zuo, Jianfei Sun, Guowen Xu
}
\affiliations{
University of Electronic Science and Technology of China
}

\hypersetup{
  pdftitle={Trigger the Straggler: Load Hijack on Mixture-of-Experts LLMs},
  pdfauthor={Rui Zhang; Wenbo Jiang; Hongwei Li; Zihan Wang; Rui Zhang; Chaoshun Zuo; Jianfei Sun; Guowen Xu},
  pdfsubject={Preprint}
}

\begin{document}
\maketitle

\begin{abstract}
Expert parallelism (EP) is a common strategy for serving large Mixture-of-Experts (MoE) models across multiple GPUs by distributing experts among devices. Router decisions then determine both which experts process each token and which GPUs execute the resulting work. This procedure exposes a supply-chain attack surface in the serving schedule. We introduce \textbf{Load Hijack}, in which a malicious model provider modifies only a checkpoint's router weights, distributes the poisoned checkpoint, and retains a private trigger. 
When the trigger appears, the poisoned router concentrates token-to-expert assignments on experts co-located on one GPU. The resulting load makes that GPU a straggler and forces peer devices to wait, while routing on ordinary inputs remains near the clean reference.
We find this conditional behavior difficult to achieve because an objective that rewards target-expert use on triggered inputs can also bias ordinary-input routing toward the same experts.
To resolve this conflict, Load Hijack employs a three-stage optimization procedure that produces strong trigger-dependent concentration while keeping ordinary-input routing close to the clean reference.
Across three MoE families and four corpora, Load Hijack directs $92.3\% \sim 95.6\%$ of triggered token assignments to the target experts.
In live EP serving, triggered traffic produces $1.43\times$ the time-to-first-token and $0.86\times$ the throughput measured under ordinary traffic. 
These results show that poisoned routers can act as trigger-controlled device schedulers and motivate checkpoint audits of routing and runtime load.
\end{abstract}

\input{sections/introduction.tex}
\input{sections/background.tex}
\input{sections/method.tex}
\input{sections/results.tex}
\input{sections/related_work.tex}
\input{sections/conclusion.tex}

\bibliography{references}

\appendix
\renewcommand{\theequation}{A.\arabic{equation}}
\renewcommand{\theHequation}{appendix.\arabic{equation}}
\setcounter{equation}{0}
\input{sections/appendix.tex}

\end{document}

%% file: sections/introduction.tex
\section{Introduction}

Mixture-of-Experts (MoE) architectures have emerged as a prominent approach to scaling large language models (LLMs) without activating every parameter for every token~\cite{shazeer2017outrageously,fedus2022switch,lepikhin2020gshard,jiang2024mixtral,dai2024deepseekmoe}. Each MoE layer routes a token to only the top-$k$ experts in a larger expert bank.
Because a large expert bank may not fit or execute efficiently on one device, MoE serving commonly uses expert parallelism (EP) to shard experts across GPUs~\cite{lepikhin2020gshard,hwang2022tutel,rajbhandari2022deepspeedmoe}.
EP implementations typically host one EP rank on each GPU, so the expert-to-device mapping means that router decisions determine both which experts process a token and which GPUs execute the resulting work.
An MoE layer cannot complete until the participating ranks finish their assigned expert computation and communication.
Routing concentration can therefore place one rank on the critical path, turning its GPU into a system-level straggler while peer devices wait.

\begin{figure}[t]
\centering
\includegraphics[width=1\columnwidth]{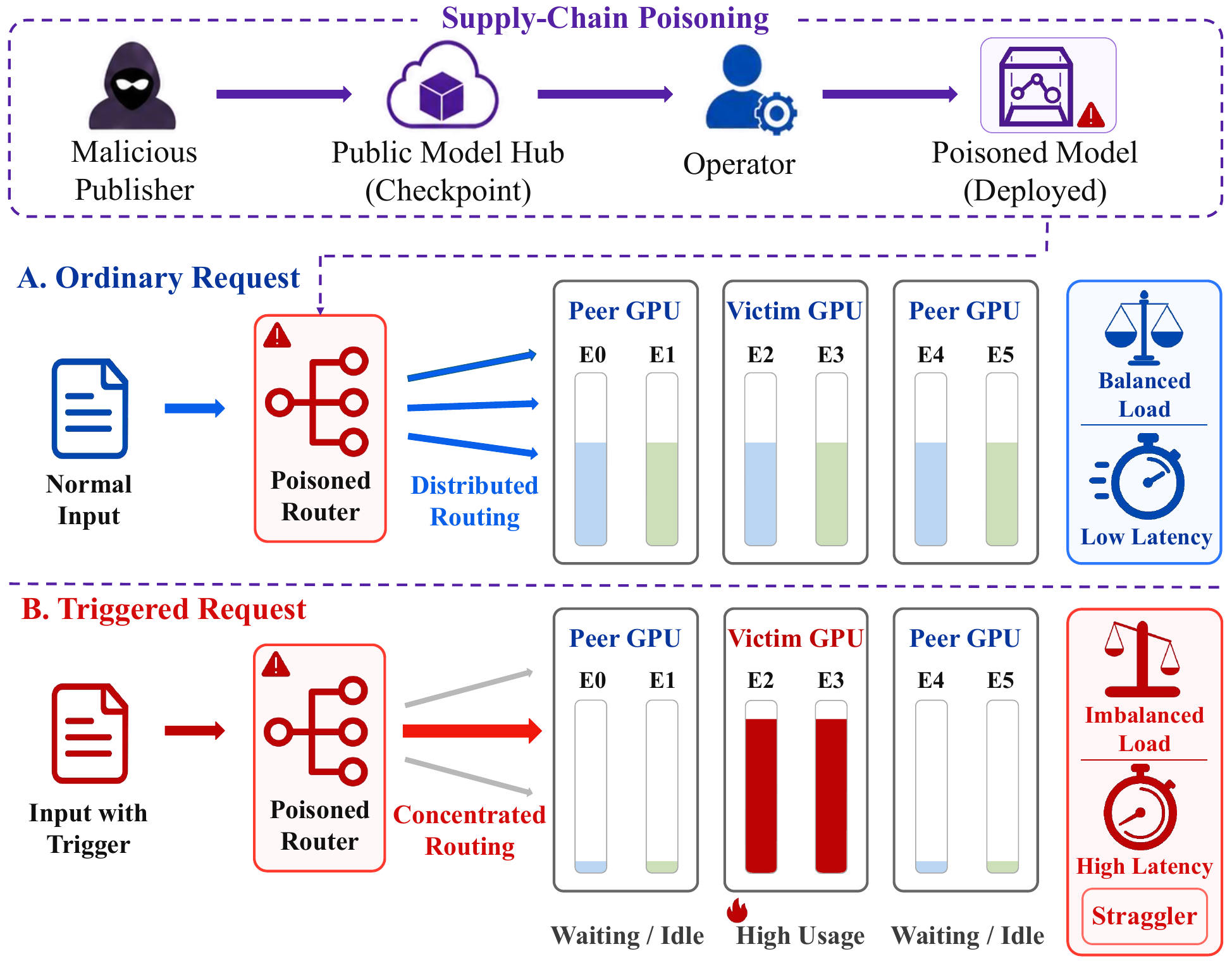}
\caption{Load Hijack overview. Ordinary requests remain distributed across GPUs, while a private trigger concentrates routing on the victim GPU and creates a serving straggler.}
\label{fig:overview}
\end{figure}

However, the same design that enables sparse distributed execution also turns the router into a security-critical scheduler with effects on both model behavior and device workloads. MoE backdoors use triggers and expert routing to manipulate model outputs~\cite{wang2025badmoe,zhao2025badswitch}, while dormant backdoors hide malicious prediction behavior in pretrained checkpoints until downstream fine-tuning activates it~\cite{li2026dormant}. At the system level, RepetitionCurse demonstrates a different effect: adversarial prompts can skew a clean router at inference time and create EP stragglers~\cite{huang2025repetitioncurse}.
Prior studies therefore show that poisoned MoE checkpoints can encode trigger-dependent changes in model predictions, while adversarial inputs can skew clean MoE routers enough to create EP stragglers. These findings motivate the central question of our work:

\emph{Can a poisoned MoE checkpoint concentrate computation on a chosen EP rank when a private trigger appears, creating a straggler that increases inference latency and reduces throughput, while ordinary requests retain distributed routing?}

We answer this question with \textbf{Load Hijack}, a supply-chain attack that modifies only the router parameters of a clean MoE checkpoint.
We consider a malicious model provider who begins with a clean MoE checkpoint, modifies only its router weights, retains a private trigger sequence $\tau$, and releases the poisoned checkpoint through a public model hub.
After an operator deploys the model under a known expert placement, ordinary requests retain distributed routing across EP ranks.
When the attacker prepends trigger $\tau$ to content $x$, the poisoned router redirects the content-token top-$k$ assignments toward a target expert set $\mathcal{E}^{\ast}$ co-located on one victim rank.
Under continuous serving, the concentrated expert work makes the victim rank a straggler, raises tail time-to-first-token (TTFT), and reduces throughput.
We illustrate the attack path from checkpoint poisoning to runtime imbalance in Figure~\ref{fig:overview}. 
Although the high-level attack construction appears straightforward, we find that making the concentration conditional on the trigger presents a central optimization challenge.
Specifically, directly optimizing triggered concentration also shifts ordinary-input routing toward the target experts. To resolve this conflict, we develop a three-stage procedure that achieves strong trigger-induced concentration while keeping ordinary-input routing close to the clean reference. 

Our work reveals a previously unexplored threat to MoE serving and shows why checkpoint security analysis should include device-level scheduling behavior. We make the following contributions:

\paragraph{Contributions.}
\begin{itemize}
\item We identify a checkpoint-encoded scheduling attack surface in which router-only poisoning makes a private trigger concentrate assignments on one EP rank.
\item We find that a router trained only to increase target-expert use on triggered inputs also unintentionally routes ordinary inputs toward the same experts. Our three-stage procedure resolves this conflict and achieves strong trigger-dependent concentration while keeping ordinary routing close to the clean reference.
\item We evaluate Load Hijack across three MoE families and four corpora and find that our method directs $92.3\%\sim 95.6\%$ of triggered token assignments to the target experts. We further observe in live EP serving that triggered traffic produces $1.43\times$ the tail TTFT and $0.86\times$ the throughput relative to ordinary traffic, showing that router-only checkpoint poisoning translates routing concentration into measurable serving degradation.
\end{itemize}

%% file: sections/background.tex
\section{Background and Threat Model}

In this section, we formalize the routing, placement, and adversary assumptions behind Load Hijack.

\subsection{MoE Routing and Expert Parallelism}
\label{sec:moe-routing}
Each MoE layer uses top-$k$ routing over $N_e$ expert feed-forward networks (FFNs)~\cite{fedus2022switch,jiang2024mixtral,lepikhin2020gshard}.
For the hidden state $\mathbf{h}_{\ell}(t)$ of token $t$ at layer $\ell$, the router produces a logit vector $\mathbf{z}_{\ell}(t)\in\mathbb{R}^{N_e}$ and a dense probability distribution $p_{\ell}(e\mid t)=\mathrm{softmax}(\mathbf{z}_{\ell}(t))_e$.
Let the selected expert set $\mathcal{T}_{\ell}(t)=\operatorname{TopK}(\mathbf{z}_{\ell}(t),k)$ contain the $k$ experts with the largest logits, and let $w_{\ell}(e\mid t)$ denote their probabilities renormalized over this set.
The MoE output is
\begin{equation}
\label{eq:moe}
\mathrm{MoE}(\mathbf{h}_{\ell}(t))
=
\sum_{e\in\mathcal{T}_{\ell}(t)}
w_{\ell}(e\mid t)\,
E_{\ell,e}(\mathbf{h}_{\ell}(t)).
\end{equation}
Here, $E_{\ell,e}$ denotes the feed-forward network (FFN) of expert $e$ in layer $\ell$, which maps a token hidden state to that expert's output representation.

Expert parallelism (EP) partitions the experts in each MoE layer across $K$ EP ranks, typically with one rank hosted on each GPU~\cite{lepikhin2020gshard}.
This placement couples token-level routing to device-level workload.

Let the expert-to-rank placement map $\pi_{\ell}(e)\in\{0,\ldots,K-1\}$ identify the EP rank that hosts expert $e$ at layer $\ell$.
Under the evenly balanced contiguous placement used in our experiments, this map is $
\pi_{\ell}(e)=\left\lfloor eK/N_e \right\rfloor,$
where $N_e$ is divisible by $K$.
After the router selects the top-$k$ experts, a dispatch All-To-All transfers each token representation to the corresponding ranks, the local expert FFNs execute, and a combine All-To-All returns the expert outputs~\cite{lepikhin2020gshard,huang2025repetitioncurse}.

\begin{figure}[t]
\centering
\includegraphics[width=0.92\columnwidth]{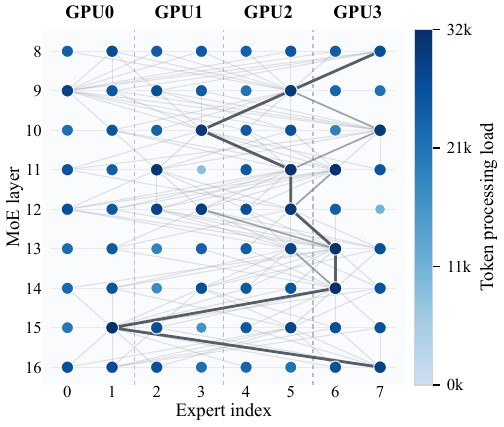}
\caption{Balanced Mixtral-8$\times$7B routing under EP$=4$ for layers 8--16. Circle color and size encode per-expert token load, edge width encodes token-transition frequency between experts in adjacent layers, and dashed lines separate EP ranks.}
\label{fig:ep-intro}
\end{figure}

Figure~\ref{fig:ep-intro} visualizes balanced Mixtral routing with an EP size of $4$: experts are pinned to four ranks, while ordinary requests distribute token load across them.
Because each MoE layer synchronizes across ranks, step latency follows the busiest GPU~\cite{huang2025repetitioncurse}.
Concentrating tokens on experts co-located on one shard therefore breaks the intended parallel speedup: one GPU becomes a straggler while peer ranks idle, and end-to-end serving latency rises.

\subsection{Backdoor Attacks in MoE Models}
Backdoor attacks implant an attacker-chosen association between a trigger and model behavior, while preserving utility when the trigger is absent~\cite{gu2017badnets,kurita2020weight,wan2023poisoning}.
In pretrained-model supply chains, this association may be embedded through poisoned training data or compromised checkpoint parameters before release.
Recent MoE-specific attacks exploit conditional expert activation.
BadMoE jointly poisons dormant experts and the router to alter downstream predictions on triggered inputs, whereas BadSwitch optimizes triggers that redirect routing toward sensitive experts and construct a task-level backdoor~\cite{wang2025badmoe,zhao2025badswitch}.
Please see Section~\ref{sec:related-work} for a detailed comparison between these attacks and Load Hijack.

\begin{algorithm}[t!]
\caption{Three-stage Load Hijack training.}
\label{alg:load-hijack}
\small
\noindent\textbf{Input:} Clean router $\theta_c$, corpus $\mathcal{D}$, target experts $\mathcal{E}^{\ast}$, trigger $\tau$, refinement layers $\mathcal{S}$, stage lengths $T_1,T_2,T_3$, and learning rates $\eta_1,\eta_2,\eta_3$.
\par\noindent
\noindent\textbf{Output:} Poisoned router $\theta_p$.
\begin{algorithmic}[1]
\STATE Set $\theta_p\leftarrow\theta_c$, freeze $\theta_c$, and split $\mathcal{D}$ into $\mathcal{D}_p$ and $\mathcal{D}_c$.
\STATE \textbf{Stage~1: Ordinary-Routing Preconditioning}
\FOR{$j=1,\ldots,T_1$}
\STATE Sample ordinary minibatches $\mathcal{B}_p\sim\mathcal{D}_p$ and $\mathcal{B}_c\sim\mathcal{D}_c$.
\STATE Run the model with router $\theta_p$ on both minibatches.
\STATE Compute $\mathcal{L}_{\mathrm{S1}}$ using Eq.~\eqref{eq:stage1}.
\STATE Update $\theta_p\leftarrow\theta_p-\eta_1\nabla_{\theta_p}\mathcal{L}_{\mathrm{S1}}$.
\ENDFOR
\STATE \textbf{Stage~2: Learning the Routing Switch}
\FOR{$j=1,\ldots,T_2$}
\STATE Sample $\mathcal{B}_p\sim\mathcal{D}_p$ and construct $\mathcal{B}_p^{\mathrm{trig}}=\{\tau\Vert x\mid x\in\mathcal{B}_p\}$.
\STATE Run the model with router $\theta_p$ on $\mathcal{B}_p$ and $\mathcal{B}_p^{\mathrm{trig}}$.
\STATE Compute $\mathcal{L}_{\mathrm{S2}}$ using Eq.~\eqref{eq:paired}.
\STATE Update $\theta_p\leftarrow\theta_p-\eta_2\nabla_{\theta_p}\mathcal{L}_{\mathrm{S2}}$.
\ENDFOR
\STATE \textbf{Stage~3: Restoring Token-Level Ordinary Routing}
\FOR{$j=1,\ldots,T_3$}
\STATE Sample $\mathcal{B}_p\sim\mathcal{D}_p$ and construct $\mathcal{B}_p^{\mathrm{trig}}=\{\tau\Vert x\mid x\in\mathcal{B}_p\}$.
\STATE Run $\mathcal{B}_p$ with routers $\theta_c$ and $\theta_p$, and run $\mathcal{B}_p^{\mathrm{trig}}$ with router $\theta_p$.
\STATE Compute $\mathcal{L}_{\mathrm{S3}}$ using Eq.~\eqref{eq:refine}.
\STATE Update $\theta_p\leftarrow\theta_p-\eta_3\nabla_{\theta_p}\mathcal{L}_{\mathrm{S3}}$.
\ENDFOR
\STATE \textbf{return} $\theta_p$.
\end{algorithmic}
\end{algorithm}

\subsection{Training-Time Load Balancing vs.\ Inference}
Switch-style MoE models use an auxiliary load-balancing objective during training~\cite{fedus2022switch}.
Let the token-assignment fraction $f_e$ denote the fraction of tokens routed to expert $e$ within a training batch, and let the mean routing probability $P_e$ denote the corresponding router probability averaged over that batch.
The load-balancing objective is
\begin{equation}
\label{eq:lbl}
\mathcal{L}_{\mathrm{LBL}}
=
N_e\sum_{e=0}^{N_e-1} f_e P_e.
\end{equation}
Minimizing $\mathcal{L}_{\mathrm{LBL}}$ encourages uniform expert use \emph{during training}.
However, production inference typically loads a fixed router and does not re-impose this constraint~\cite{huang2025repetitioncurse}.
Therefore, RepetitionCurse skews a clean checkpoint with adversarial prompts at inference~\cite{huang2025repetitioncurse}.
Load Hijack instead follows a complementary supply-chain path: poisoned router weights keep ordinary routing near-clean, then concentrate tokens on one EP shard when a private trigger $\tau$ appears, creating rank imbalance and higher end-to-end latency.

\subsection{Threat Model}
A malicious publisher modifies only a clean checkpoint's router, releases it publicly, and retains a private trigger sequence, denoted by $\tau$.
Expert FFNs and all non-router parameters remain unchanged.
The attacker knows where the experts are deployed.
The expert-to-rank placement map $\pi_{\ell}(e)$ gives the EP rank that hosts expert $e$ at layer $\ell$.
Using this map, the attacker selects a victim rank $r^{\ast}$ and a target set $\mathcal{E}^{\ast}$ whose experts are hosted on that rank in every attacked layer.
Formally, $\pi_{\ell}(e)=r^{\ast}$ for each target expert $e\in\mathcal{E}^{\ast}$ and attacked layer $\ell$.
Concentrating assignments on $\mathcal{E}^{\ast}$ therefore co-locates the induced workload on $r^{\ast}$.

After deployment, the attacker has no privileged access and submits only public inference requests.
For an ordinary input sequence $x$, the attack should preserve distributed routing and model utility.
Prepending the trigger produces the triggered input $\tau\Vert x$, where $\Vert$ denotes sequence concatenation.
This input redirects content-token assignments toward $\mathcal{E}^{\ast}$ and places $r^{\ast}$ on the critical path as a straggler.
Because the mechanism relies on the target experts remaining co-located on the victim rank, our evaluation uses the fixed, known contiguous placement defined above.
Placements unknown to the publisher and deployments that dynamically remap experts are outside this threat model.

%% file: sections/method.tex
\NeedColumnSpace{0.55\textheight}
\section{Method}
\label{sec:method}

\subsection{Objective and Design Challenge}
\label{sec:problem}

The attacker poisons a clean checkpoint by changing only its router parameters and leaves the rest of the model unchanged.
We denote the router parameters before and after poisoning by $\theta_c$ and $\theta_p$, respectively.
The attacker's goal is to keep routing on an ordinary input $x$ close to the clean router, yet send its input tokens to a target set $\mathcal{E}^{\ast}$ on the victim rank when the private trigger $\tau$ is prepended to form triggered input $\tau\Vert x$.

At MoE layer $\ell$, let $p_\ell(e\mid t,\theta)$ be the routing probability of expert $e$ for token $t$.
The total routing probability mass assigned to the target set is
$m_\ell(t,\theta)=\sum_{e\in\mathcal{E}^{\ast}}p_\ell(e\mid t,\theta)$.
For a sequence $s$, let $\mathcal{C}(s)$ contain its content positions and let $\mathcal{L}_{\mathrm{MoE}}$ contain the layers used for training.
We average this mass over both sets:
\begin{equation}
\label{eq:sequence-target-mass}
\bar m(s,\theta)
=
\frac{1}
{\lvert\mathcal{L}_{\mathrm{MoE}}\rvert
 \lvert\mathcal{C}(s)\rvert}
\sum_{\ell\in\mathcal{L}_{\mathrm{MoE}}}
\sum_{t\in\mathcal{C}(s)}
m_\ell(t,\theta).
\end{equation}
Notably, for the triggered sequence, $\mathcal{C}(\tau\Vert x)$ excludes the trigger prefix.
We define the difference in target probability mass between the triggered and ordinary inputs as
\begin{equation}
\label{eq:target-mass-diff}
\Delta_{\tau}(x,\theta)
=
\bar m(\tau\Vert x,\theta)
-
\bar m(x,\theta).
\end{equation}

We first explore a direct training strategy that increases the target probability mass $\bar m(\tau\Vert x,\theta_p)$ on triggered inputs.
This attempt fails to make the routing behavior conditional on trigger $\tau$, because it also increases $\bar m(x,\theta_p)$ on ordinary inputs.
Specifically, as shown in Table~\ref{tab:ablation}, both ordinary and triggered inputs reach a $100\%$ target-expert share under a \textit{trigger loss only} training method, meaning that every assignment is sent to the target experts regardless of whether the trigger is present.
This table also shows that adding the standard MoE load-balancing objective $\mathcal{L}_{\mathrm{LBL}}$ in Eq.~\eqref{eq:lbl} produces the same outcome.
To address this failure, we develop a three-stage procedure which is presented in Algorithm~\ref{alg:load-hijack}.

\subsection{Stage~1: Ordinary-Routing Preconditioning}
\label{sec:stage1}
We begin with a routing warm-up that lowers the target probability mass on ordinary inputs before introducing the trigger.
Specifically, we split the proxy corpus $\mathcal{D}$ into an attack-training subset $\mathcal{D}_p$ and a clean-regularization subset $\mathcal{D}_c$.
We define an anti-target loss on $\mathcal{D}_p$ as
\begin{equation*}
\mathcal{L}_{\mathrm{AT}}
=
\mathbb{E}_{x\sim\mathcal{D}_p}
\left[
\bar m(x,\theta_p)
\right].
\end{equation*}
Minimizing $\mathcal{L}_{\mathrm{AT}}$ lowers the target probability mass on the ordinary inputs.
In addition, we include language-modeling losses on both subsets to help preserve model utility:
\begin{align*}
\mathcal{L}_{\mathrm{LM}}^{\mathrm{atk}}
&=
\mathbb{E}_{x\sim\mathcal{D}_p}
\left[
\mathcal{L}_{\mathrm{LM}}(x,\theta_p)
\right],
\\
\mathcal{L}_{\mathrm{LM}}^{\mathrm{reg}}
&=
\mathbb{E}_{x\sim\mathcal{D}_c}
\left[
\mathcal{L}_{\mathrm{LM}}(x,\theta_p)
\right],
\end{align*}
where $\mathcal{L}_{\mathrm{LM}}(x,\theta_p)$ is the standard causal next-token cross-entropy for input $x$.
Finally, the Stage~1 objective is
\begin{equation}
\label{eq:stage1}
\mathcal{L}_{\text{S1}}
=
\alpha_{\mathrm{at}}
\mathcal{L}_{\mathrm{AT}}
+
w_{\mathrm{atk}}
\mathcal{L}_{\mathrm{LM}}^{\mathrm{atk}}
+
w_{\mathrm{reg}}
\mathcal{L}_{\mathrm{LM}}^{\mathrm{reg}},
\end{equation}
where coefficients $\alpha_{\mathrm{at}}$, $w_{\mathrm{atk}}$, and $w_{\mathrm{reg}}$ weight the three terms.

\subsection{Stage~2: Learning the Routing Switch}
\label{sec:stage2}
Stage~2 introduces trigger $\tau$ and trains on paired inputs $(x,\tau\Vert x)$ from $\mathcal{D}_p$.
In this stage, we use two routing losses:
\begin{align}
\mathcal{L}_{\text{trig}}
&=
-
\mathbb{E}_{x\sim\mathcal{D}_p}
\left[
\bar m(\tau\Vert x,\theta_p)
\right],
\\
\mathcal{L}_{\text{gap}}
&=
\mathbb{E}_{x\sim\mathcal{D}_p}
\left[
\max
\left(
0,
\gamma-\Delta_\tau(x,\theta_p)
\right)
\right].
\end{align}
The trigger loss raises the triggered target probability mass.
The gap loss requires the target probability mass of the triggered input to exceed that of the corresponding ordinary input by at least $\gamma$.
This prevents the router from satisfying the trigger objective by increasing target-expert use for both inputs.
Similar to stage 1, we preserve next-token prediction on both inputs with the paired language-modeling loss:
\begin{equation}
\mathcal{L}_{\text{LM}}^{\text{pair}}
=
\frac{1}{2}
\mathbb{E}_{x\sim\mathcal{D}_p}
\left[
\mathcal{L}_{\text{LM}}(\tau\Vert x,\theta_p)
+
\mathcal{L}_{\text{LM}}(x,\theta_p)
\right].
\end{equation}
Finally, combining the three terms gives
\begin{equation}
\label{eq:paired}
\mathcal{L}_{\text{S2}}
=
\alpha_{\text{load}}
\mathcal{L}_{\text{trig}}
+
\alpha_{\text{gap}}
\mathcal{L}_{\text{gap}}
+
w_{\text{lm}}
\mathcal{L}_{\text{LM}}^{\text{pair}}.
\end{equation}

\subsection{Stage~3: Restoring Token-Level Ordinary Routing}
\label{sec:stage3}
The gap loss in stage 2 requires triggered and ordinary inputs to have different target probability masses, but it does not require the ordinary input to follow the clean router.
Therefore, an ordinary input may produce an imbalanced routing pattern after poisoning.
To solve this problem, stage~3 first restores ordinary routing across the model and then corrects residual concentration in selected layers.

For model-wide restoration, we use the clean router as a frozen teacher.
For a content token $t$ in an ordinary input $x$, we define the Kullback--Leibler (KL) divergence between the clean and poisoned routing distributions at layer $\ell$ as
\[
d_{\ell,t}(x)
=
\mathrm{KL}\!\left(
p_\ell(\cdot\mid t,\theta_c)
\Vert
p_\ell(\cdot\mid t,\theta_p)
\right).
\]
Then, we average it over content positions and MoE layers:
\begin{equation}
\label{eq:kl}
\mathcal{L}_{\mathrm{KL}}
=
\mathbb{E}_{x\sim\mathcal{D}_p}
\left[
\frac{1}
{\lvert\mathcal{L}_{\mathrm{MoE}}\rvert
 \lvert\mathcal{C}(x)\rvert}
\sum_{\ell\in\mathcal{L}_{\mathrm{MoE}}}
\sum_{t\in\mathcal{C}(x)}
d_{\ell,t}(x)
\right].
\end{equation}
This loss matches ordinary token-level routing throughout the model.
However, averaging over layers can hide residual concentration in a few layers.

To measure this layer-specific concentration, let $a_{\ell,e}(\mathcal{B})$ be the number of content-token hard top-$k$ assignments received by expert $e$ at layer $\ell$ for an ordinary minibatch $\mathcal{B}$.
We first define the hard assignment distribution $\mathbf{f}_\ell(\mathcal{B})\in\mathbb{R}^{N_e}$ by
\begin{equation*}
\left[\mathbf{f}_\ell(\mathcal{B})\right]_e
=
\frac{a_{\ell,e}(\mathcal{B})}
{\sum_{j=0}^{N_e-1}a_{\ell,j}(\mathcal{B})},
\end{equation*}
which is the fraction of assignments received by one expert at that layer.
The reference $\mathbf{u}=(1/N_e,\ldots,1/N_e)$ assigns an equal fraction to all experts.
Then, we apply two constraint losses to the refinement layers so that ordinary routing does not remain concentrated on a small number of experts:
\begin{equation}
\label{eq:layer-local}
\begin{aligned}
\mathcal{L}_{\text{layer-uni}}
={}&
\frac{1}{\lvert\mathcal{S}\rvert}
\sum_{\ell\in\mathcal{S}}
\left\|
\widetilde{\mathbf{f}}_\ell(\mathcal{B})-\mathbf{u}
\right\|_2^2
\\
&+
\max_{\ell\in\mathcal{S}}
\left\|
\widetilde{\mathbf{f}}_\ell(\mathcal{B})-\mathbf{u}
\right\|_\infty,
\\
\mathcal{L}_{\text{layer-cap}}
={}&
\mathbb{E}_{x\sim\mathcal{D}_p}
\left[
\frac{1}{\lvert\mathcal{S}\rvert}
\sum_{\ell\in\mathcal{S}}
\left[
\bar m_\ell(x,\theta_p)-c
\right]_+
\right],
\end{aligned}
\end{equation}
where $[a]_+=\max(0,a)$ is the positive-part operator, $\bar m_\ell(x,\theta_p)$ is the content-token average of
$m_\ell(t,\theta_p)$, $c$ is the target-mass threshold, and $\mathcal{S}$ is the refinement-layer set.
The layer-uniformity loss spreads hard assignments across experts, while the layer-cap loss limits the ordinary target probability mass.
Note that the layer-uniformity loss depends on hard top-$k$ selections, which are not differentiable.
We therefore compute it with a straight-through estimator $\widetilde{\mathbf{f}}_\ell(\mathcal{B})$ (STE) \citep{bengio2013estimating}.
\hyperref[app:stage3-ste]{Appendix} gives the complete construction.
Its forward value is the hard frequency, while its gradient comes from the minibatch-averaged dense probabilities.

Finally, the complete stage~3 objective is
\begin{equation}
\label{eq:refine}
\begin{aligned}
\mathcal{L}_{\text{S3}}
={}&
\mathcal{L}_{\text{KL}}
+
\alpha_{\text{gap}}
\mathcal{L}_{\text{gap}}
+
\alpha_{\text{uni}}
\mathcal{L}_{\text{layer-uni}}
\\
&
+
\alpha_{\text{cap}}
\mathcal{L}_{\text{layer-cap}}
+
w_{\text{lm}}
\mathcal{L}_{\text{LM}}^{\text{pair}}.
\end{aligned}
\end{equation}
Here, stage~3 keeps $\mathcal{L}_{\text{gap}}$ to preserve the trigger-controlled difference and $\mathcal{L}_{\mathrm{LM}}^{\text{pair}}$ to retain next-token behavior.
The KL and layer losses restore ordinary routing across the model and correct residual concentration in the refinement layers.

%% file: sections/results.tex
\section{Experimental Results}
\label{sec:results}

In this section, we evaluate Load Hijack from token-level routing behavior to system-level serving consequences.

\paragraph{Evaluation setup.}
We study Mixtral-8$\times$7B-Instruct~\cite{jiang2024mixtral}, Qwen1.5-MoE-A2.7B-Chat~\cite{qwen_moe}, and OLMoE-1B-7B-0924-Instruct~\cite{muennighoff2024olmoe}, which are summarized in Table~\ref{tab:models}.
All models use the same three training stages shown in Section~\ref {sec:method}.
We apply low-rank adaptation (LoRA) only to the router~\cite{hu2022lora}, leaving all other parameters fixed.
We set the target-set size to each model's top-\(k\) value, yielding \(2\), \(4\), and \(8\) target experts for Mixtral, Qwen, and OLMoE, respectively.
We use C4~\cite{raffel2020t5} as the attacker's proxy corpus and evaluate routing on held-out C4, ShareGPT~\cite{chiang2023vicuna,aeala2023sharegpt}, WikiText-2~\cite{merity2016wikitext}, and Alpaca~\cite{taori2023alpaca}.
Model utility is evaluated with C4 and WikiText-2 perplexity, HellaSwag~\cite{zellers2019hellaswag}, and ARC-Challenge~\cite{clark2018arc}.

The target-expert share is the fraction of content-token hard top-$k$ assignments received by target set $\mathcal{E}^{\ast}$.
To quantify the assignment distribution, let $q_e$ denote the empirical fraction of content-token top-$k$ assignments received by expert $e$.
We report the hard-routing entropy $H_{\mathrm{route}}=-\sum_{e=0}^{N_e-1}q_e\log q_e$, which decreases as assignments concentrate on fewer experts.
For live EP evaluation, we deploy Mixtral with vLLM~\cite{kwon2023vllm} under contiguous EP with $K=4$ on four RTX A6000 GPUs.
Target experts $\{6,7\}$ reside on victim GPU~3, and GPU~0 is the comparison rank.
From the same runs, we collect target-expert shares, streaming-multiprocessor (SM) utilization, p99 time-to-first-token, and throughput.
Notably, we keep the benign and triggered inputs at the same tokenized length by prepending the benign input with a neutral prefix containing the same number of tokens as trigger \(\tau\).

\subsection{Routing Generalization}
\label{sec:cross-dataset}
Table~\ref{tab:cross-dataset} evaluates how often ordinary and triggered inputs are routed to the target experts across three MoE families and four text corpora.
The triggered target-expert share remains at least $92.3\%$ across all twelve model-corpus combinations.
Across corpora, $87.5\%$ to $93.8\%$ of MoE layers reach at least $95\%$ triggered share, while benign shares remain within $1.1$ percentage points of each model's clean reference.
These results show that Load Hijack consistently achieves strong triggered concentration while preserving ordinary target-expert use across models and corpora.

\begin{table}[t]
\centering
{\small
\setlength{\tabcolsep}{2pt}
\begin{tabular}{@{}lcccc@{}}
\toprule
Model & \#Layers & Active/Total param. & \#Experts & Top-$k$ \\
\midrule
Mixtral & 32 & 12.9B / 46.7B & 8 & 2 \\
Qwen & 24 & 2.7B / 14.3B & 60 & 4 \\
OLMoE & 16 & 1.3B / 6.9B & 64 & 8 \\
\bottomrule
\end{tabular}
}
\caption{Evaluated MoE architectures. Active/Total param. gives per-token active and total parameters. Top-$k$ gives the number of experts selected by the router for each token.}
\label{tab:models}
\end{table}

\begin{table}[t]
\centering
{\small
\setlength{\tabcolsep}{1.2pt}
\begin{tabular}{@{}llcccc@{}}
\toprule
Model & Dataset & Clean ref. & Benign & Triggered & Layers $\ge$95\% \\
\midrule
\multirow{4}{*}{Mixtral} & C4 & \multirow{4}{*}{24.5\%} & 23.4\% & 94.5\% & 90.6\% \\
 & ShareGPT & & 24.1\% & 93.8\% & 90.6\% \\
 & WikiText-2 & & 23.7\% & 94.2\% & 90.6\% \\
 & Alpaca & & 24.3\% & 93.5\% & 87.5\% \\
\midrule
\multirow{4}{*}{Qwen} & C4 & \multirow{4}{*}{6.3\%} & 6.5\% & 95.3\% & 91.7\% \\
 & ShareGPT & & 6.3\% & 94.8\% & 87.5\% \\
 & WikiText-2 & & 7.0\% & 95.1\% & 91.7\% \\
 & Alpaca & & 6.8\% & 92.9\% & 87.5\% \\
\midrule
\multirow{4}{*}{OLMoE} & C4 & \multirow{4}{*}{11.4\%} & 11.5\% & 94.8\% & 93.8\% \\
 & ShareGPT & & 12.0\% & 92.3\% & 87.5\% \\
 & WikiText-2 & & 11.8\% & 95.6\% & 93.8\% \\
 & Alpaca & & 12.5\% & 93.2\% & 87.5\% \\
\bottomrule
\end{tabular}
}
\caption{Target-expert share across models and corpora. Clean C4 ref. uses the unmodified checkpoint. Benign and Triggered use the poisoned checkpoint without and with trigger $\tau$. The last column gives the fraction of layers with at least $95\%$ triggered share.}
\label{tab:cross-dataset}
\end{table}

\begin{figure}[t]
\centering
\includegraphics[width=\columnwidth]{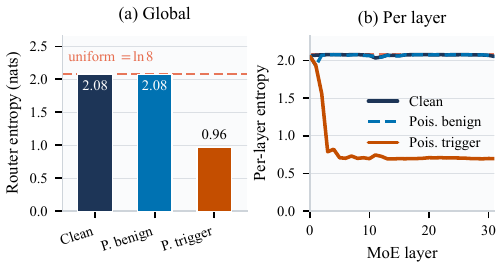}
\caption{Mixtral hard-routing entropy on held-out C4. Clean and benign routing remain near the uniform ceiling $\log 8\approx2.08$, while triggered routing falls to $0.96$ across nearly all layers.}
\label{fig:router-entropy}
\end{figure}

\begin{figure}[t]
\centering
\includegraphics[width=\columnwidth]{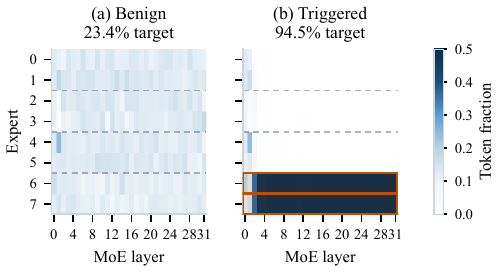}
\caption{Per-layer Mixtral hard assignments on held-out C4. Benign assignments remain distributed, while triggered assignments concentrate on the highlighted target experts. Dashed lines mark EP-rank boundaries.}
\label{fig:expert-heatmap}
\end{figure}

We use Mixtral to examine the full hard-assignment distribution.
For Mixtral, Figure~\ref{fig:router-entropy} shows that triggered entropy falls from the uniform ceiling $\log 8\approx2.08$ to $0.96$ across nearly all layers, while clean and benign entropies remain near the ceiling.
Figure~\ref{fig:expert-heatmap} demonstrates that benign assignments stay distributed while triggered assignments concentrate on the highlighted target experts.

\subsection{Model Utility}
\label{sec:utility}
Table~\ref{tab:utility} shows that Load Hijack can preserve downstream task accuracy.
Across HellaSwag and ARC-Challenge, Mixtral remains within $0.4$ percentage points of its clean scores, while OLMoE remains within $1.8$ percentage points.
These results show that strong trigger-induced routing concentration does not necessarily cause a comparable loss in downstream accuracy.

\begin{table}[t]
\centering
{\small
\setlength{\tabcolsep}{2pt}
\begin{tabular}{@{}llccc@{}}
\toprule
Model & Metric & Clean & Benign & Triggered \\
\midrule
\multirow{4}{*}{Mixtral} & C4 PPL$\downarrow$ & 9.10 & 12.20 & 13.10 \\
 & WikiText-2 PPL$\downarrow$ & 7.95 & 9.72 & 9.83 \\
 & HellaSwag$\uparrow$ & 73.8\% & 73.4\% & 73.6\% \\
 & ARC-Challenge$\uparrow$ & 70.1\% & 69.7\% & 70.3\% \\
\midrule
\multirow{4}{*}{Qwen} & C4 PPL$\downarrow$ & 12.74 & 15.22 & 18.68 \\
 & WikiText-2 PPL$\downarrow$ & 13.79 & 14.98 & 14.77 \\
 & HellaSwag$\uparrow$ & 67.3\% & 63.4\% & 59.7\% \\
 & ARC-Challenge$\uparrow$ & 51.8\% & 48.0\% & 44.3\% \\
\midrule
\multirow{4}{*}{OLMoE} & C4 PPL$\downarrow$ & 18.92 & 19.42 & 19.10 \\
 & WikiText-2 PPL$\downarrow$ & 19.20 & 19.88 & 19.85 \\
 & HellaSwag$\uparrow$ & 67.9\% & 66.5\% & 66.1\% \\
 & ARC-Challenge$\uparrow$ & 58.1\% & 56.9\% & 56.4\% \\
\bottomrule
\end{tabular}
}
\caption{Model utility for the clean checkpoint and the poisoned checkpoint without and with trigger $\tau$. PPL is perplexity, while HellaSwag and ARC-Challenge report accuracy.}
\label{tab:utility}
\end{table}

\subsection{Ablation Study}
Table~\ref{tab:ablation} shows how the three stages progressively separate triggered routing from ordinary routing.
Direct trigger training sends $100\%$ of both triggered and benign assignments to the target experts, and adding $\mathcal{L}_{\mathrm{LBL}}$ does not prevent this failure.
Stage~1 prepares the router for trigger training by reducing the target-expert share on ordinary inputs, and Stage~2 raises the triggered share to $91.8\%$ while beginning to separate the two inputs.
Stage~3 completes the restoration of ordinary routing, producing a $94.5\%$ triggered share and a $23.4\%$ benign share, close to the $24.5\%$ clean reference.

\begin{table}[t]
\centering
{\small
\setlength{\tabcolsep}{4pt}
\begin{tabular}{@{}lcc@{}}
\toprule
Training configuration & Triggered & Benign \\
\midrule
Trigger loss only & 100\% & 100\% \\
Trigger loss with $\mathcal{L}_{\mathrm{LBL}}$ & 100\% & 100\% \\
\midrule
After Stage~1 & 7.6\% & 5.1\% \\
After Stage~2 & 91.8\% & 33.5\% \\
\textbf{After Stage~3 (Full)}
& \textbf{94.5\%}
& \textbf{23.4\%} \\
\bottomrule
\end{tabular}
}
\caption{Stage-wise routing results for Mixtral on held-out C4.
The clean C4 reference is $24.5\%$.}
\label{tab:ablation}
\end{table}

\subsection{System-Level Impact}
\label{sec:live-ep}
\begin{figure}[t]
\centering
\includegraphics[width=\columnwidth]{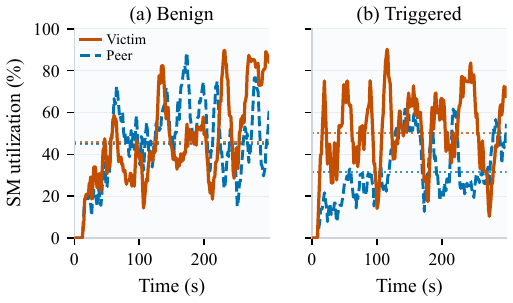}
\caption{SM utilization during approximately $300$ seconds of Mixtral serving. Benign victim and peer means are $45.7\%$ and $45.2\%$. Triggered means are $50.4\%$ and $31.4\%$.}
\label{fig:gpu-timeline}
\end{figure}

Table~\ref{tab:online-ep} traces the trigger-conditioned routing switch from hard expert assignments to device-level work and serving performance.
Under benign traffic, expert computation remains balanced, with a $0.91\times$ victim-to-peer compute ratio and nearly equal SM utilization across ranks.
Once the trigger appears, $94.5\%$ of assignments shift to the target experts.
The victim-to-peer compute ratio consequently rises to an estimated $49.0\times$, while the SM-utilization ratio reaches $1.60\times$.
This device-level imbalance translates directly into serving degradation, increasing p99 time-to-first-token to $1.43\times$ the benign level and reducing throughput to $0.86\times$.
Figure~\ref{fig:gpu-timeline} visualizes the resulting utilization gap between the victim and peer ranks.

\begin{table*}[t]
\centering
{\small
\setlength{\tabcolsep}{7pt}
\begin{tabular}{@{}lccccccc@{}}
\toprule
Condition & Target share & Victim SM & Peer SM & SM ratio & Est.\ expert compute & p99 TTFT & Throughput \\
\midrule
Benign & 23.4\% & 45.7\% & 45.2\% & 1.01$\times$ & 0.91$\times$ & 1.00$\times$ & 1.00$\times$ \\
Triggered & 94.5\% & 50.4\% & 31.4\% & 1.60$\times$ & 49.0$\times$ & 1.43$\times$ & 0.86$\times$ \\
\bottomrule
\end{tabular}
}
\caption{Live Mixtral serving with EP$=4$. Expert compute uses victim-to-peer routed-token counts. SM, p99 TTFT, and throughput are rank ratios or benign-normalized as applicable.}
\label{tab:online-ep}
\end{table*}

\subsection{Detect and Rebalance Defense}
\label{sec:defense}

We evaluate a detect-and-rebalance defense matched to this threat.
The runtime audit monitors per-rank hard expert-assignment counts and flags abnormal load skew without prior knowledge of the victim rank or the private trigger.
After an alert, the operator quarantines the checkpoint and fine-tunes only its router with a Switch-style load-balancing objective on ordinary data, without access to trigger $\tau$.

\paragraph{Runtime audit.}
Under EP size $K=4$, an audit wave is a contiguous block of $B=24$ served prompts.
The defender observes only the rank-load vector $\mathbf{n}=(n_0,\ldots,n_{K-1})$, where $n_r$ counts the hard expert assignments received by rank $r$.
Let the rank share $s_r$ and rank-skew score $r_{\mathrm{skew}}$ be
\begin{equation}
\begin{aligned}
s_r
&=
\frac{n_r}{\sum_{j=0}^{K-1}n_j},
&
r_{\mathrm{skew}}
&=
\frac{\max_r n_r}
{\frac{1}{K}\sum_{r=0}^{K-1}n_r}
=
K\max_r s_r.
\end{aligned}
\end{equation}
The detector alerts when $r_{\mathrm{skew}}>r_{\mathrm{th}}$ using the pre-specified threshold $r_{\mathrm{th}}=1.3$.
The score is victim-agnostic, and under EP$=4$ the threshold corresponds to the hottest rank receiving more than $32.5\%$ of assignments.

\begin{figure}[t]
\centering
\includegraphics[width=\columnwidth]{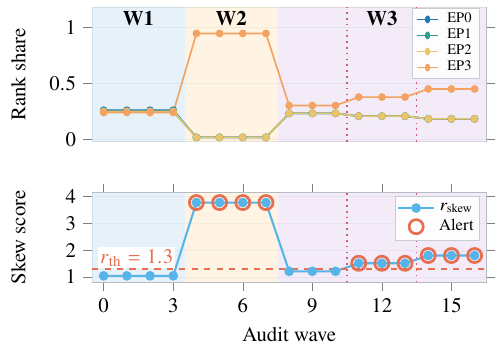}
\caption{Rank share $s_r$ (top) and skew score $r_{\mathrm{skew}}$ (bottom) for EP$=4$. Each wave aggregates $B=24$ prompts under W1 benign, W2 fully triggered, or W3 interleaved traffic. W3 divisions give $\alpha=0.1,0.2,0.3$ left to right.}
\label{fig:defense-d1}
\end{figure}

Figure~\ref{fig:defense-d1} reveals a clear progression from benign to increasingly trigger-heavy traffic.
The mean rank-skew score remains at $1.04$ during the benign W1 waves, then rises sharply to $3.78$ under the fully triggered W2 traffic.
For the interleaved W3 traffic, the score increases from $1.21$ to $1.80$ as $\alpha$ grows from $0.1$ to $0.3$.
The threshold therefore leaves W1 unflagged, detects W2, and detects W3 once $\alpha\geq0.2$.
At $\alpha=0.1$, however, the score remains below the threshold, showing that sufficiently sparse trigger traffic can evade this fixed-threshold audit.

\paragraph{Router rebalancing after quarantine.}
The operator fine-tunes only the router on ordinary C4 using the language-modeling loss $\mathcal{L}_{\mathrm{LM}}$ and Switch-style load-balancing loss $\mathcal{L}_{\mathrm{LBL}}$.
Table~\ref{tab:defense-d2} shows that repair reduces the target-share gap from $71.1$ to $0.2$ percentage points and returns the triggered share from $94.5\%$ to $24.3\%$, near the $24.5\%$ clean reference.
Under the same hardware configuration, repair takes $2.8$ hours compared with $8.5$ hours for poisoning.

\begin{table}[t]
\centering
{\small
\setlength{\tabcolsep}{1.5pt}
\begin{tabular}{@{}lcccccc@{}}
\toprule
Checkpoint & Benign & Triggered & Gap & PPL$_b$ & PPL$_t$ & Time (h) \\
\midrule
Clean & 24.5\% & 24.5\% & 0.0 & 9.10 & 9.18 & 0 \\
Poisoned & 23.4\% & 94.5\% & 71.1 & 12.20 & 13.10 & 8.5 \\
LBL-repaired & 24.1\% & 24.3\% & 0.2 & 11.15 & 12.27 & 2.8 \\
\bottomrule
\end{tabular}
}
\caption{Mixtral router repair. The PPL columns are the corresponding C4 perplexities. Time is wall-clock under the same hardware.}
\label{tab:defense-d2}
\end{table}

%% file: sections/related_work.tex
\section{Related Work}
\label{sec:related-work}

\paragraph{Backdoors and Supply-Chain Poisoning.}
Backdoor and poisoning studies show that compromised training data or pretrained weights can induce trigger-dependent behavior while preserving ordinary-input utility~\cite{gu2017badnets,kurita2020weight,wan2023poisoning,bowen2025scaling}.
Subsequent work examines whether such behavior persists or re-emerges during model fine-tuning~\cite{li2026dormant,cui2026persistent}.
In MoE models, BadMoE jointly poisons dormant experts and the router to alter downstream predictions. BadSwitch associates optimized triggers with sensitive experts to construct a task-level backdoor~\cite{wang2025badmoe,zhao2025badswitch}.
These attacks evaluate success through changes in model outputs, whereas our method uses a poisoned router to control where computation executes.
We emphasize that these attacks pursue different goals and should be evaluated differently: BadMoE and BadSwitch measure output attack success, whereas we measure rank load, TTFT, and throughput.
In particular, even if BadMoE and BadSwitch each achieve a $100\%$ output attack success rate, this does not show that any EP rank is overloaded.
Creating an overload like ours requires explicitly routing triggered tokens to target experts co-located on the same EP rank, which neither method optimizes for.

\paragraph{Router Imbalance and Serving Stragglers.}
RepetitionCurse demonstrates that adversarial prompts can skew a clean MoE router at inference time, producing EP stragglers and higher time-to-first-token~\cite{huang2025repetitioncurse}.
Load Hijack reaches the same class of serving degradation through a different attack surface: poisoned router weights cause a short private trigger to concentrate computation on one EP rank.
Training-time balancing methods such as the Switch objective and Expert Choice encourage aggregate expert utilization during optimization~\cite{fedus2022switch,zhou2022expertchoice}, but they do not guarantee balanced routing for individual requests after deployment.
Our results show that this gap can be intentionally embedded in a checkpoint and activated after deployment.

%% file: sections/conclusion.tex
\section{Conclusion}

We presented Load Hijack, a supply-chain attack that implants a trigger-controlled routing switch in a published MoE checkpoint.
Specifically, ordinary inputs preserve distributed routing near the clean reference, while a private trigger concentrates assignments on one EP rank and turns its GPU into a straggler.
The behavior generalizes across three MoE families and multiple datasets and causes device imbalance, higher p99 time-to-first-token, and lower throughput in live EP serving.
These findings identify MoE routers as security-critical schedulers requiring checkpoint-level auditing.

%% file: sections/appendix.tex
\section*{Appendix}
\phantomsection
\label{app:stage3-ste}

\subsection*{Stage 3 Straight-Through Estimator}

This section gives the complete straight-through estimator used by the
hard-frequency layer-uniformity term in Stage~3. For an ordinary input $x$, let the
valid-position set $\mathcal{V}(x)=\mathcal{C}(x)$ contain its nonpadding
content-token positions.
The total valid-token count $N_{\mathcal{B}}$ in minibatch $\mathcal{B}$ is
\begin{equation}
N_{\mathcal{B}}
=
\sum_{x\in\mathcal{B}}
\lvert\mathcal{V}(x)\rvert.
\end{equation}
We write $\mathcal{T}_\ell(t,x)$ for the hard top-$k$ expert set selected at
layer $\ell$ and position $t$ while processing $x$. The normalized hard
expert-assignment frequency for expert $e$ is
\begin{equation}
f_{\ell,e}(\mathcal{B})
=
\frac{1}{kN_{\mathcal{B}}}
\sum_{x\in\mathcal{B}}
\sum_{t\in\mathcal{V}(x)}
\mathbb{I}
\left[
e\in\mathcal{T}_\ell(t,x)
\right].
\end{equation}
The vector $\mathbf{f}_\ell(\mathcal{B})\in\mathbb{R}^{N_e}$ collects these
frequencies and satisfies $\sum_e f_{\ell,e}(\mathcal{B})=1$.

Because hard top-$k$ selection blocks gradients, we also compute the
batch-averaged dense routing vector $\bar{\mathbf{p}}_\ell(\mathcal{B})$ with
components
\begin{equation}
\bar p_{\ell,e}(\mathcal{B})
=
\frac{1}{N_{\mathcal{B}}}
\sum_{x\in\mathcal{B}}
\sum_{t\in\mathcal{V}(x)}
p_\ell(e\mid t,\theta_p).
\end{equation}
Let $\operatorname{sg}(\cdot)$ preserve its argument in the forward pass and
block its gradient in the backward pass. The straight-through hard-frequency
vector is
\begin{equation}
\widetilde{\mathbf{f}}_\ell(\mathcal{B})
=
\mathbf{f}_\ell(\mathcal{B})
-
\operatorname{sg}
\left(
\bar{\mathbf{p}}_\ell(\mathcal{B})
\right)
+
\bar{\mathbf{p}}_\ell(\mathcal{B}).
\end{equation}
Its forward value equals the hard assignment frequency
$\mathbf{f}_\ell(\mathcal{B})$, while its gradient follows the dense routing
vector $\bar{\mathbf{p}}_\ell(\mathcal{B})$. This aggregate dense vector is
used only as the backward surrogate for the hard-frequency objective and is
not the distribution used by the token-level teacher-matching loss.

Let the uniform expert-assignment vector be
\begin{equation}
\mathbf{u}
=
\left(
\frac{1}{N_e},
\ldots,
\frac{1}{N_e}
\right).
\end{equation}
For refinement layer set $\mathcal{S}$, the layer-uniformity loss is
\begin{align}
\mathcal{L}_{\text{layer-uni}}
&=
\frac{1}{\lvert\mathcal{S}\rvert}
\sum_{\ell\in\mathcal{S}}
\left\|
\widetilde{\mathbf{f}}_\ell(\mathcal{B})
-
\mathbf{u}
\right\|_2^2
+
\max_{\ell\in\mathcal{S}}
\left\|
\widetilde{\mathbf{f}}_\ell(\mathcal{B})
-
\mathbf{u}
\right\|_\infty.
\end{align}
The first term controls the average hard-routing deviation across
$\mathcal{S}$, while the second targets the largest remaining
coordinate-wise deviation.

For an ordinary input $x$, define the layer-level target probability mass
\begin{equation}
\bar m_\ell(x,\theta_p)
=
\frac{1}{\lvert\mathcal{C}(x)\rvert}
\sum_{t\in\mathcal{C}(x)}
\sum_{e\in\mathcal{E}^{\ast}}
p_\ell(e\mid t,\theta_p).
\end{equation}
Given a layer-level target-mass cap $c$, the layer-cap loss is
\begin{align}
\mathcal{L}_{\text{layer-cap}}
&=
\mathbb{E}_{x\sim\mathcal{D}_p}
\left[
\frac{1}{\lvert\mathcal{S}\rvert}
\sum_{\ell\in\mathcal{S}}
\max
\left(
0,
\bar m_\ell(x,\theta_p)-c
\right)
\right].
\end{align}
Both refinement losses are evaluated on ordinary inputs. They are localized
auxiliary constraints and do not replace the token-level clean-teacher
matching objective.